\documentstyle[12pt,twoside]{article}
\def\greaterthansquiggle{\raise.3ex\hbox{$>$\kern-.75em\lower1ex\hbox{$\sim$}}}

\def\lessthansquiggle{\raise.3ex\hbox{$<$\kern-.75em\lower1ex\hbox{$\sim$}}}

\newcommand{\beq}{\begin{equation}}
\newcommand{\eeq}{\end{equation}}
\newcommand{\beqa}{\begin{eqnarray}}
\newcommand{\eeqa}{\end{eqnarray}}
\newcommand{\beqan}{\begin{eqnarray*}}
\newcommand{\eeqan}{\end{eqnarray*}}
\newcommand{\ba}{\begin{array}}
\newcommand{\ea}{\end{array}}
\newcommand{\no}{\nonumber}

\def\nz{\ifmmode {I\hskip -3pt N} \else {\hbox {$I\hskip -3pt N$}}\fi}
\def\zz{\ifmmode {Z\hskip -4.8pt Z} \else
       {\hbox {$Z\hskip -4.8pt Z$}}\fi}
\def\qz{\ifmmode {Q\hskip -5.0pt\vrule height6.0pt depth 0pt
       \hskip 6pt} \else {\hbox
       {$Q\hskip -5.0pt\vrule height6.0pt depth 0pt\hskip 6pt$}}\fi}
\def\rz{\ifmmode {I\hskip -3pt R} \else {\hbox {$I\hskip -3pt R$}}\fi}
\def\cz{\ifmmode {C\hskip -4.8pt\vrule height5.8pt\hskip 6.3pt} \else
       {\hbox {$C\hskip -4.8pt\vrule height5.8pt\hskip 6.3pt$}}\fi}
\def\au{{\setbox0=\hbox{\lower1.36775ex%
\hbox{''}\kern-.05em}\dp0=.36775ex\hskip0pt\box0}}
\def\ao{{}\kern-.10em\hbox{``}}

\renewcommand{\baselinestretch}{1.5} 
\normalsize
\begin{document}
\bibliographystyle{plain}

\begin{titlepage}
\begin{flushright}
UWThPh-2002-31\\

\end{flushright}
\vspace{2cm}
\begin{center}
{\Large \bf Comment on the Decoupling of  UV and IR \\ Divergencies
within Dimensional
Regularization\\[13pt] in Noncommutative Theories}\\[40pt]

Helmuth H\"uffel* \\
Institut f\"ur Theoretische Physik \\
Universit\"at Wien \\
Boltzmanngasse 5, A-1090 Vienna, Austria
\vfill

{\bf Abstract}
\end{center}
\renewcommand{\baselinestretch}{1.0} 
\small

In a recent paper \cite{nar} it has been claimed that to one-loop 
order in  
noncommutative  $\phi ^4$ scalar field theory using 
dimensional regularization the UV and IR divergencies decouple. 
We point out that this statement is incorrect. 

\vfill
\begin{enumerate}
\item[*)] Email: helmuth.hueffel@univie.ac.at

\end{enumerate}
\end{titlepage}
\renewcommand{\baselinestretch}{1.5} 
\normalsize
The  mixing  of the ultraviolet and infrared divergencies of nonplanar 
diagrams constitutes
one of the most serious problems for the
renormalizability of noncommutative quantum field theories
\cite{Minwalla,VanRaam,Chepelev}.
Specifically we consider the noncommutative Euclidean scalar 
$\phi^4$ theory leaving aside  discussions on the Minkowskian 
space-time formulation \cite{freden,sib1,sib2} as well as omitting 
gauge theory complications, see e.g. \cite{raimar2}. The Lagrangian 
is given by
\beq
{\cal L}(x)={1\over 2}(\partial_\mu\phi)^2+{1\over 2}m^2\phi^2+{g^2\over 4!}
\phi\star\phi\star\phi\star\phi~
\eeq
where the $\star$ product is defined as
\beq
(\phi_1\star\phi_2)(x)=e^{{i\over 2}
\theta_{\mu\nu}\partial^y_\mu\partial^z_\nu}\phi_1(y)\phi_2(z)|_{y=z=x}~
\eeq
and  where $\theta_{\mu\nu}$ is the constant, antisymmetric 
noncommutativity parameter matrix.

Several authors   used  dimensional regularization 
for analyzing divergencies
within the noncommutative setting of quantum field theory, see e.g. 
\cite{Aref1,Aref2,brandt,sad} 
and  also recently  \cite{nar}. In this last work it was asserted 
that  to one-loop order the UV/IR 
mixing should be seen as a regularization prescription depending 
artifact and that using dimensional regularization the UV and IR 
divergencies are decoupling. We point out 
that this statement is incorrect:

At the one loop level the essential contribution  of 
the nonplanar tadpole diagram 
is  given  
by the n-dimensional integral
\begin{equation}
I(n,p) =  \int {d^{n}k\over (2\pi)^{n}}\,
{e^{i k_\mu\theta^{\mu\nu}p_\nu}\over (k^{2}+M^{2})}=\frac{m^{\,\frac{-2
+ n}{2}}\,
    (p \circ p)^{\,\frac{2 - n}{4}}\,
    K_{\frac{-2 + n}{2}}\,(
     m\,{\sqrt{p\circ p}})}{{\left(
       2\,\pi  \right) }^{\frac{n}{2}}}
\end{equation}
where $p{\circ} p\equiv -p^\mu \theta^2_{\mu\nu}p^\nu.$ 
As in $n=4$ 
dimensions the integral is not absolutely convergent, the limits 
$n \rightarrow 4$ and $p_{\mu} \rightarrow 0$ cannot be expected to 
commute. 
Performing first the $n \rightarrow 4$ limit (i.e. considering 
vanishing dimensional regularization) we immediately 
have
\beq
I(4,p):=\lim_{n \to 4}\,I(n,p)=\frac{m\,K_{1}(
     m\,{\sqrt{{p \circ p}}})}{4\,
    {\pi }^2\,{\sqrt{{p \circ p}}}}.
\eeq
Subsequently we obtain 
for small $p_{\mu}$ 
\beq
I(4,p)\Longrightarrow\frac{1}{4\,{\pi }^2\,{p \circ p}} + 
\frac{m^2 (-1+2 \gamma + log \frac{m^2 \,p \circ p}{4})}{16 \pi^2}+
  {O(p \circ p)}\quad {\rm for}\quad  p_{\mu} \to 0.
  \eeq
Conversely, keeping the ultraviolet 
regularization by fixing
 $n < 2$  (allowing analytic continuation only later on)  
 the limit $p_{\mu} \rightarrow 0$  
can be performed easily exploiting basic properties of the Bessel functions
\beqa
I(n,0)&:=&\lim_{p_{\mu} \to 0}\,I(n,p)=\lim_{p_{\mu} \to 0}\,\left[{(p
\circ p)}^{-\frac{n}{2}}\left (-\frac{ {\pi }^
           {1 - \frac{n}{2}}\,
          \csc (\frac{n\,\pi }{2}) }{4\,
        {\Gamma}(2 - 
          \frac{n}{2})}\,
          p \circ p  + 
     {O(p \circ p)}^2
     \right)\right.\no\\
	 &&\mbox{}\left. + 
  \frac{m^{-2 + n}\,
      {\pi }^{1 - \frac{n}{2}}\,
      \csc (\frac{n\,\pi }{2})}{2^n\,
      {\Gamma}(\frac{n}{2})} + 
   {O(p \circ p)}^1\right]=(4 \pi)^{-\frac{n}{2}} m^{-2+n} 
\Gamma (1-\frac{n}{2})
\eeqa
Eventually, we obtain for  $n \rightarrow 4$ 
\beq
I(n,0)\Longrightarrow\frac{m^2}{8\,{\pi }^2\,(n-4) } + 
  {O(1)} \quad {\rm for}\quad  {\rm n} \to 4.
\eeq
The discrepancy of (5) and (7) is a manifestation of  UV/IR mixing 
in complete analogy to the discussion 
 of \cite{Minwalla,VanRaam,Chepelev}.  We conclude that  to one-loop 
 order   there does not 
 exist 
UV/IR decoupling within the  dimensional regularization 
scheme.

\section*{Acknowledgements}

I am grateful for valuable discussions with H. Grosse and R. Wulkenhaar.


\begin{thebibliography}{99}
	
\bibitem{nar} S. Narison, ``Decoupling of the UV and IR divergences
within dimensional
regularization in non-commutative theories and some phenomenological 
implications'', hep-ph/0208225.

	
\bibitem{Minwalla}
S.~Minwalla, M.~Van Raamsdonk and N.~Seiberg,
``Noncommutative perturbative dynamics'',
JHEP {\bf 0002} (2000) 020.
\bibitem{VanRaam}
M.~Van Raamsdonk and N.~Seiberg,
``Comments on noncommutative perturbative dynamics'',
JHEP {\bf 0003} (2000) 035.

\bibitem{Chepelev}
I.~Chepelev and R.~Roiban,
``Convergence theorem for non-commutative Feynman graphs and  renormalization'',
JHEP {\bf 0103} (2001) 001.


\bibitem{freden} D. Bahns, S. Doplicher, K. Fredenhagen and G. 
Piacitelli, ``On the unitarity problem in space-time noncommutative 
theories'', Phys. Lett. {\bf B533} (2002) 178.

\bibitem{sib1} Y. Liao and K. Sibold, ``Time ordered perturbation 
theory on noncommutative space-time: Basic rules'', hep-th/0205269. 

\bibitem{sib2} Y. Liao and K. Sibold, ``Time ordered perturbation 
theory on noncommutative space-time. 2. Unitarity'', hep-th/0206011.

\bibitem{raimar2} R. Wulkenhaar, ``Quantum field theories on 
noncommutative R**4 versus \mbox{theta-expanded} quantum field
theories'', 
hep-th/0206018.



\bibitem{Aref1}
I.~Aref'eva, D.~Belov and A.~Koshelev,
``Two-loop diagrams in noncommutative $\phi^4_4$ theory'', 
Phys.\ Lett.\ B {\bf 476}, 431 (2000);\,``A note on UV/IR for
noncommutative complex scalar field'',
hep-th/0001215.

\bibitem{Aref2}
I.~Aref'eva, D.~Belov, A.~Koshelev and O.~Rychkov, 
``Renormalizability and UV/IR mixing in noncommutative theories with
scalar fields,''
Phys.\ Lett.\ B {\bf 487}, 357 (2000);\,
``UV/IR mixing for noncommutative complex scalar field theory.  II:
Interaction with gauge fields,''
Nucl.\ Phys.\ Proc.\ Suppl.\  {\bf 102}, 11 (2001).


\bibitem{brandt} F. Brandt, A. Das and J. Frenkel, ``General 
structure of the photon selfenergy in noncommutative QED'', Phys.Rev. 
{\bf D65} (2002) 85017.

\bibitem{sad} N. Sadooghi, M. Mohammadi, ``On the beta function and 
conformal anomaly of noncommutative QED with adjoint matter fields,
hep-th/0206137. 
\end{thebibliography}
\end{document}